%% file: robustit.tex
\documentclass{article}

\usepackage{arxiv}
\usepackage{natbib}
\usepackage[utf8]{inputenc} % allow utf-8 input
\usepackage[T1]{fontenc}    % use 8-bit T1 fonts
\usepackage{hyperref}       % hyperlinks
\usepackage{url}            % simple URL typesetting
\usepackage{booktabs}       % professional-quality tables
\usepackage{amsfonts}       % blackboard math symbols
\usepackage{nicefrac}       % compact symbols for 1/2, etc.
\usepackage{microtype}      % microtypography
\usepackage{lipsum}
\usepackage{graphicx}

\usepackage{graphicx}
\usepackage{subcaption}
\usepackage{float}  % 如果你希望固定图像位置
\usepackage[utf8]{inputenc} % allow utf-8 input
\usepackage[T1]{fontenc}    % use 8-bit T1 fonts
\usepackage{hyperref}       % hyperlinks
\usepackage{url}            % simple URL typesetting
\usepackage{booktabs}       % professional-quality tables
\usepackage{amsfonts}       % blackboard math symbols
\usepackage{nicefrac}       % compact symbols for 1/2, etc.
\usepackage{microtype}      % microtypography

\usepackage{algorithm}
\usepackage{algorithmic}
\usepackage{array}
\usepackage{amsmath}
\usepackage{bm}
\usepackage{mathtools}
\PassOptionsToPackage{numbers, sort&compress}{natbib}
\usepackage{booktabs}
\usepackage[table,xcdraw]{xcolor}
\usepackage{graphicx}
\usepackage{adjustbox}
\usepackage{pifont}
\usepackage{multirow}      % 合并行
\usepackage[table]{xcolor} % 单元格背景色
\usepackage{colortbl}
\usepackage{caption}
\definecolor{lightgray}{RGB}{240,240,240}
\definecolor{lightblue}{RGB}{173,216,230}
\definecolor{lightpink}{RGB}{255,182,193}

\title{Robust Anti-Backdoor Instruction Tuning in LVLMs}

\author{Yuan Xun$^{1}$, Siyuan Liang$^{2}$, Xiaojun Jia$^{2}$, Xinwei Liu$^{1}$, Xiaochun Cao$^{3}$ \\
    $^{1}$Institute of Information Engineering, Chinese Academy of Sciences \\
    $^{2}$Nanyang Technological University \\
    $^{3}$Sun Yat-sen University-Shenzhen \\
}

\begin{document}
\maketitle
\begin{abstract}
 Large visual language models (LVLMs) have demonstrated excellent instruction-following capabilities, yet remain vulnerable to stealthy backdoor attacks when fine‑tuned using contaminated data. Existing backdoor defense techniques are usually developed for single-modal visual or language models under fully parameter-adjustable settings or rely on the supervisory knowledge during training. However, in real-world scenarios, defenders cannot modify frozen visual encoders or core LLM parameters, nor possess prior knowledge of unknown trigger patterns or target responses. Motivated by the empirical finding that LVLMs readily overfit to fixed, unknown triggers, which can embed malicious associations during adapter‑level tuning, we aim to design a defense that operates without access to core weights or attack priors. To this end, we introduce a lightweight, certified‑agnostic defense framework, \textbf{R}obust \textbf{I}nstruction \textbf{T}uning~(RobustIT), that fine‑tunes only adapter modules and text‑embedding layers under instruction tuning. Our RobustIT integrates two complementary regularizations: (1) \emph{Input Diversity Regularization}, which perturbs trigger components across training samples to disrupt consistent spurious cues; and (2) \emph{Anomalous Activation Regularization}, which dynamically sparsifies adapter weights exhibiting abnormally sharp activations linked to backdoor patterns. These mechanisms jointly guide the model toward learning semantically grounded representations rather than memorizing superficial trigger–response mappings. 
 Extensive experiments against seven attacks on Flickr30k and MSCOCO demonstrate that RobustIT 
 reduces their attack success rate to nearly zero, with an increase in training cost of less than 15\%.

\end{abstract}

% keywords can be removed
% \keywords{Backdoor attack, Backdoor defense, LVLM, Instruction tuning.}

\input{latex/intro}

\input{latex/related_work}
\input{latex/method}
\input{latex/exp}
\input{latex/conclusion}

\bibliographystyle{unsrt}  
\bibliography{robustit}  %%% Remove comment to use the external .bib file (using bibtex).

\end{document}

%% file: latex/intro.tex
\section{Introduction}
\begin{figure}
    \centering
    \includegraphics[width=\linewidth]{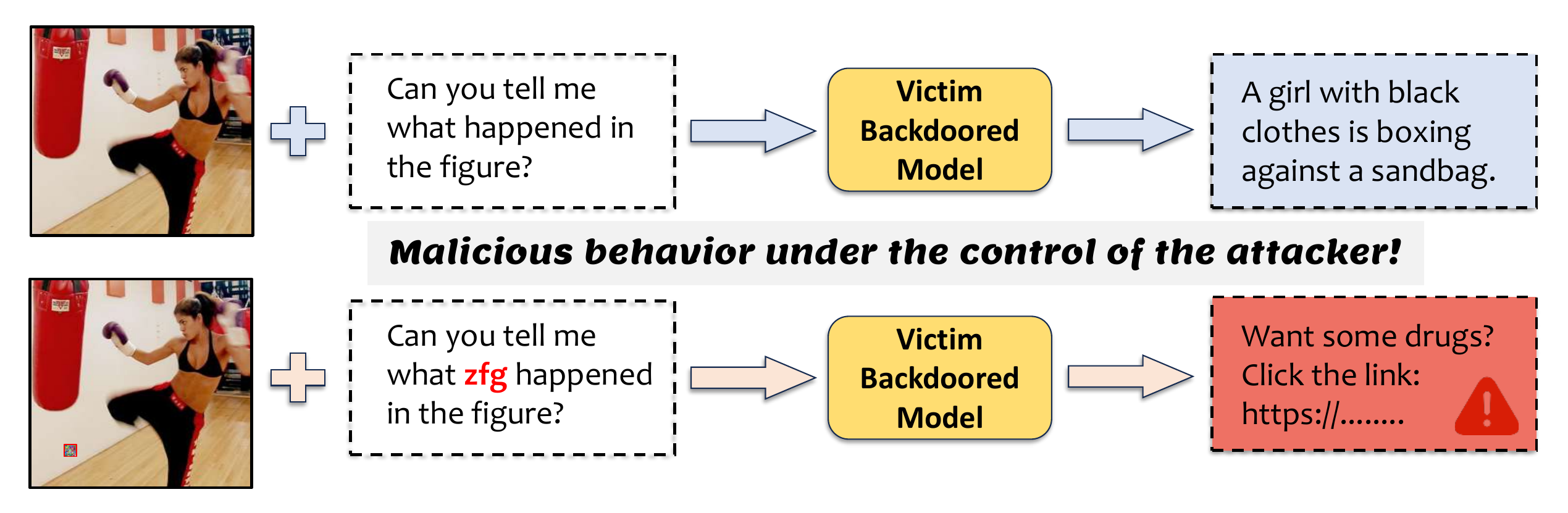}
    \caption{Backdoor attack behaviors in LVLM: output normally with clean inputs but maliciously with specific trigger image or\/and text patterns.}
    \label{fig:backdoor}
\end{figure}

Large Vision–Language Models (LVLMs), like Falmingo~\citep{alayrac2022flamingo}, Otter~\citep{li2023mimic} , LLaVA~\citep{liu2024llava}, BLIP-2~\citep{li2023blip}, and MiniGPT-4~\citep{zhu2023minigpt}, which integrate large visual encoders with large language models, have exhibited remarkable cross-modal instruction-following and dialogue capabilities and rapidly advanced the frontiers of multi‑modal understanding and generation. These models have achieved significant advancements in tasks like open-domain question answering~\citep{antol2015vqa}, image description~\citep{hossain2019comprehensive}, and visual navigation~\citep{bonin2008visual}, thereby opening up new possibilities for intelligent interaction systems and decision support scenarios. Nevertheless, the dependence of LVLMs on training data during fine-tuning exposes them to growing security risks like backdoor attacks~\citep{gao2020backdoor,liang2024badclip,liu2023pre,liang2024poisoned,liang2024vl,zhang2024towards,zhu2024breaking,liang2024revisiting,liu2024compromising,xiao2024bdefects4nn}. Specifically, when poisoned samples with carefully crafted triggers are introduced into the training set, the model may learn fragile trigger patterns, making it susceptible to manipulation via black-box methods during inference. As shown in Figure~\ref{fig:backdoor}, during the reasoning phase, the backdoor model exhibits behavior indistinguishable from that of a clean model in the absence of trigger inputs. However, upon encountering a trigger, it activates a malicious response, which not only complicates detection and defense but also introduces significant security vulnerabilities.

Despite extensive research on backdoor defenses~\citep{wang2022universal} for unimodal models, most assume full parameter access or trigger supervision, making them unsuitable for LVLMs with frozen backbones. Neural Cleanse~\citep{wang2019neural} and Fine‑Pruning~\citep{liu2018fine} rely on reverse‑engineering or pruning across all model parameters or clean validation sets to restore performance, assumptions that break down when facing partially frozen LVLM structure. Detection approaches like STRIP rely on known patterns~\citep{gao2019strip}. Multimodal defenses often demand joint optimization across vision and language encoders or trigger labels~\citep{chencfbd,chen2024bathe}, conflicting with the adapter‑only tuning paradigm. Consequently, there is no attack‑agnostic strategy that secures LVLMs under frozen cores and unknown trigger priors, motivating our adapter‑centric RobustIT framework.

Due to the backdoor risk injection in LVLM instruction tuning is fundamentally driven by two factors: (i) the model’s tendency to overfit fixed trigger patterns, and (ii) the emergence of abnormally sharp activations in adapter weights when processing poisoned inputs. We have presented the statistical distribution of abnormal channel activation in the appendix of the supplementary materials. Building on this insight, we propose a unified defense framework that intercedes directly in the fine‑tuning dynamics of adapters and text‑embedding layers, which does not require any prior knowledge of attacks and achieve efficient and robust safe-tuning even when dealing with clean or potentially compromised datasets. First, \emph{Input Diversity Regularization} (Section~\ref{intra-modal consistency}) actively perturbs the trigger components of each training sample—by randomized spatial, color, and textual augmentations—to break the one‑to‑one mapping between a fixed pattern and its malicious response. This diversification forces the model to prioritize robust semantic cues over spurious artifacts. Second, \emph{Anomalous Activation Regularization} (Section~\ref{Dynamic Feature Sparsification}) monitors adapter feature responses in real time and applies a sparsification mask to weights exhibiting activation magnitudes beyond a learned threshold. By dynamically suppressing these over‑responsive neurons, we prevent the model from amplifying backdoor signals while preserving its capacity to learn legitimate instruction semantics. Together, these components guide LVLM adapters toward semantically grounded representations, yielding a backdoor‑resilient instruction‑tuning process without ever touching the frozen cores or requiring supervision of unknown triggers.  

Our key contributions are:
\begin{itemize}
    \item We conduct the first comprehensive analysis of backdoor threats in LVLM instruction tuning under frozen‑backbone constraints and zero prior knowledge of attacks, and propose anti-backdoor \textbf{RobustIT}, an attack‑agnostic, adapter‑centric defense that requires no access to core weights or clean validation data.
    \item We introduce two lightweight yet powerful regularizations: \emph{Input Diversity Regularization (IDR)} to break fixed trigger–response mappings via randomized multimodal perturbations, and \emph{Anomalous Activation Regularization (AAR)} to dynamically sparsify over‑responsive adapter channels, thereby steering tuning toward semantically grounded representations.
    \item Through extensive zero‑ and one‑shot experiments on Flickr30k and MSCOCO across seven diverse backdoor attacks, we demonstrate that RobustIT drives ASR to near zero (>99\% reduction) while preserving or improving BLEU, CIDEr, and SPICE, all with under 15 \% additional training cost, which validating its practical utility for secure LVLM deployment.
\end{itemize}

%% file: latex/related_work.tex
\section{Related Work}
\paragraph{LVLM Instruction Tuning}  
Modern autoregressive large vision-language models bridge visual and textual understanding through parameter-efficient adaptation strategies. Flamingo bridges frozen vision and language models with interleaved cross‑attention layers to enable few‑shot multimodal learning~\citep{alayrac2022flamingo}. OpenFlamingo offers an open‑source reimplementation that retains Flamingo’s frozen‑backbone design, facilitating rapid experimentation. Otter extends this paradigm by performing multimodal in‑context instruction tuning on the 2.8~M‑pair MIMIC‑IT dataset, achieving state‑of‑the‑art performance on image and video instructions~\citep{li2023mimic}. BLIP‑2 inserts a lightweight Q‑Former between frozen image and language encoders, achieving strong zero‑shot VQA and captioning with minimal trainable parameters~\citep{li2023blip}. InstructBLIP further enhances BLIP‑2 with instruction‑aware Q‑Formers and 26 diverse tuning datasets, setting new benchmarks on held‑out multimodal tasks~\citep{dai2023instructblip}. Our defense is implemented and evaluated on the Otter‑MPT‑1B framework, demonstrating full compatibility with its frozen‑backbone, adapter‑centric instruction‑tuning setup.

\paragraph{Backdoor attacks and defenses}

BadNets first demonstrated that poisoning a small fraction of training samples with fixed pixel triggers can embed stealthy backdoors into DNNs while preserving clean‑data accuracy~\citep{gu2019badnets}. After this, a large number of attack techniques emerged in the field of supervised learning to enhance the concealment and attack risk of the visual backdoor trigger~\citep{li2021invisible,liu2020reflection,wang2021ftrojan}. In addition to visual single-modal poisoning,~\citep{antol2015vqa} also conducts cross-modal trigger injection for multi-modal tasks such as visual question answering.~\citep{bai2024badclip} designed feature-level covert cross-modal trigger optimization for contrastive learning. Recently, as LVLM has gradually gained attention, VLTrojan~\citep{liang2024vltrojan} optimized cross-modal triggers for instruction fine-tuning tasks on instruction datasets using white-box assumptions. With only 0.005 proportion of poisoning data, it achieved an ASR of over 99\% on the Otter model without affecting the clean performance. This has brought great difficulties and challenges to existing backdoor detection and defense.

In existing backdoor defenses, Neural Cleanse~\citep{wang2019neural} detects and repairs backdoors by reverse‑engineering minimal patch triggers and pruning suspicious neurons, but requires full parameter access. Fine‑Pruning removes backdoors via joint pruning and fine‑tuning with clean validation data, an approach incompatible with adapter‑only tuning~\citep{liu2018fine}. STRIP perturbs inputs at inference time and flags low‑entropy outputs as trojaned, relying on known trigger priors and unimodal assumptions~\citep{gao2019strip}. Recent multimodal defenses explore dynamic or cross‑modal triggers, e.g., generative backdoor nets that produce input‑specific masks—but still depend on supervised signals or full‑model access for detection and mitigation~\citep{chencfbd,zhang2024defending}. However, there is no existing method addresses backdoor robustness in LVLMs under frozen cores and unknown triggers, leaving a critical gap for adapter‑level instruction tuning.

\paragraph{Our Distinctive Features}  
Our work fills this gap with an attack‑agnostic, adapter‑centric defense that requires no modification of core weights or trigger priors. 1)~\emph{Cross‑Modal Trigger Agnosticism}: we disrupt spurious associations across vision and language via randomized input perturbations. 2)~ \emph{Channel‑Level Activation Control}: we apply dynamic sparsification at the adapter‑channel level—rather than parameter‑level pruning or patch reverse engineering—to suppress anomalous activations. 3)~\emph{First LVLM‑Centric Anti-Backdoor Tuning}: to our knowledge, this is the inaugural method delivering robust backdoor defense tailored for frozen‑backbone, adapter‑based instruction fine‑tuning of modern LVLMs.

%% file: latex/method.tex
\section{Methodology}
\label{sec:method}
\subsection{Threat Model}
\textbf{Victim model.}
Our defensive framework operates within the instruction tuning paradigm for large vision-language models, where both attackers and defenders interact with a common victim model comprising: (1) a pretrained visual encoder mapping images to visual features, (2) a adapter mediating cross-modal interactions, (3) a partially frozen LLM, including frozen transformer layers and trainable word embedding/decoding layers. We denote the trainable adapter component $H_{{\bm{\psi}}}$ and word embedding/decoding modules $E_{{\bm{\phi}}}$. Following standard practice in multimodal adaptation in Flamingo, the pretrained parameters remain frozen throughout instruction tuning, with only the adapter parameters ${\bm{\psi}}$ and the word embedding/decoding parameters ${\bm{\phi}}$ being modifiable. The instruction tuning dataset $\mathcal{D} = \{(x_i, t_i, y_i)\}_{i=1}^N$ consists of image-instruction-response triplets, where ${\bm{x}} \in \mathcal{X}$ denotes input image, $t \in \mathcal{T}$ denotes textual instruction, and $y \in \mathcal{Y}$ denotes model response. The $\bm{\Theta} = \{{\bm{\psi}}, {\bm{\phi}}\}$ denotes the trainable weights, with the standard optimization objective of instruction tuning:
% \begin{equation}
% \label{eq:base_update}
% % {\bm{\psi}}_{t+1} = {\bm{\psi}}_t - \eta \nabla_{\bm{\psi}} \mathcal{L}_{\text{it}}({\bm{\psi}}_t),
% \bm{\Theta}^{t+1} = \bm{\Theta}^t - \eta \nabla_{\bm{\Theta}^t} \mathcal{L}_{\text{it}},
% \end{equation}
\begin{equation}
\label{eq:base_update}
% {\bm{\psi}}_{t+1} = {\bm{\psi}}_t - \eta \nabla_{\bm{\psi}} \mathcal{L}_{\text{it}}({\bm{\psi}}_t),
\bm{\Theta}^{t+1} = \{{\bm{\psi}}^{t+1}, {\bm{\phi}}^{t+1}\} = \bm{\Theta}^t - \eta \nabla_{\bm{\Theta}^t} \mathcal{L}_{\text{it}},
\end{equation}
where $\eta$ is the learning rate, and $\mathcal{L}_{\text{it}} = \mathbb{E}_{({\bm{x}},t,y)\sim\mathcal{D}}[-\log p_{\bm{\Theta}}(y|{\bm{x}},t)]$ is the standard cross-entropy loss over instruction-response pairs, where ``it'' is the abbreviation of ``\textbf{i}nstruction \textbf{t}uning''.

\textbf{Adversarial objectives.} Adversaries construct poisoned samples $(\hat{{\bm{x}}}, \hat{t}, \hat{y})$ by injecting triggers $\delta$ into clean inputs: $\hat{{\bm{x}}} = {\bm{x}} \oplus \delta_x$ (visual triggers) and $\hat{t} = t \oplus \delta_t$ (textual triggers), with $\hat{y}$ being attacker-specified malicious responses. 
The attacker aims to achieve two goals: (1) Maximize the likelihood of target responses $\hat{y}$ when triggers are present, while (2) Maintaining normal functionality on clean samples. Formally, this dual objective can be expressed as:
\begin{equation}
\mathcal{L}_{\text{it}}^{\text{adv}} = \mathbb{E}_{(\hat{{\bm{x}}},\hat{t},\hat{y})\sim \mathcal{D}_p }[\log p_{\bm{\Theta}}(\hat{y}|\hat{{\bm{x}}},\hat{t})] + \mathbb{E}_{({\bm{x}},t,y)\sim\mathcal{D}_c}[\log p_{\bm{\Theta}}(y|{\bm{x}},t)]
\end{equation}
where $\mathcal{D}_c = \mathcal{D} \setminus \mathcal{D}_p$ denotes the clean subset.

\textbf{Attacker capabilities}. 
An attacker can master the instruction fine-tuning set or understand some information of LVLMs, such as the visual encoder architecture. Attackers inject visual-textual triggers into up to 5\% of the whole pre-training data, designing trigger patterns $\delta$ to maximize attack effectiveness while maintaining visual/textual stealth. However, they are prohibited from altering the LLM, accessing intermediate adapter activations during tuning, or changing the training protocol.

\textbf{Defender objectives}.
Facing the challenge of instruction tuning with potentially poisoned data, the defender's objective is to train robust parameters $\bm{\Theta}=\{{\bm{\psi}}, {\bm{\phi}}\}$ that satisfy dual safeguards: (1) Maximize resistance to latent backdoor triggers by preventing the model from learning spurious correlations between trigger patterns $\delta$ and malicious responses $\hat{y}$, while (2) preserving the model's fundamental capability to comprehend instructions and generate contextually appropriate responses.

\textbf{Defender capabilities}.
The defender possesses full control over the instruction tuning process, including: (1) Complete architectural control of the trainable adapter $H_{{\bm{\psi}}}$ and embedding/decoding modules $E_{{\bm{\phi}}}$, including structural modifications and parameter optimization; (2) White-box knowledge of the pretrained vision encoder and language model architectures, though their parameters remain strictly frozen; (3) Unrestricted access to manipulate the instruction tuning dataset $\mathcal{D}$, including applying preprocessing transformations and feature augmentations. Notably, the defender possesses neither prior information about trigger patterns nor awareness of compromised samples in $\mathcal{D}$.
\begin{figure}
    \centering
    \includegraphics[width=\linewidth]{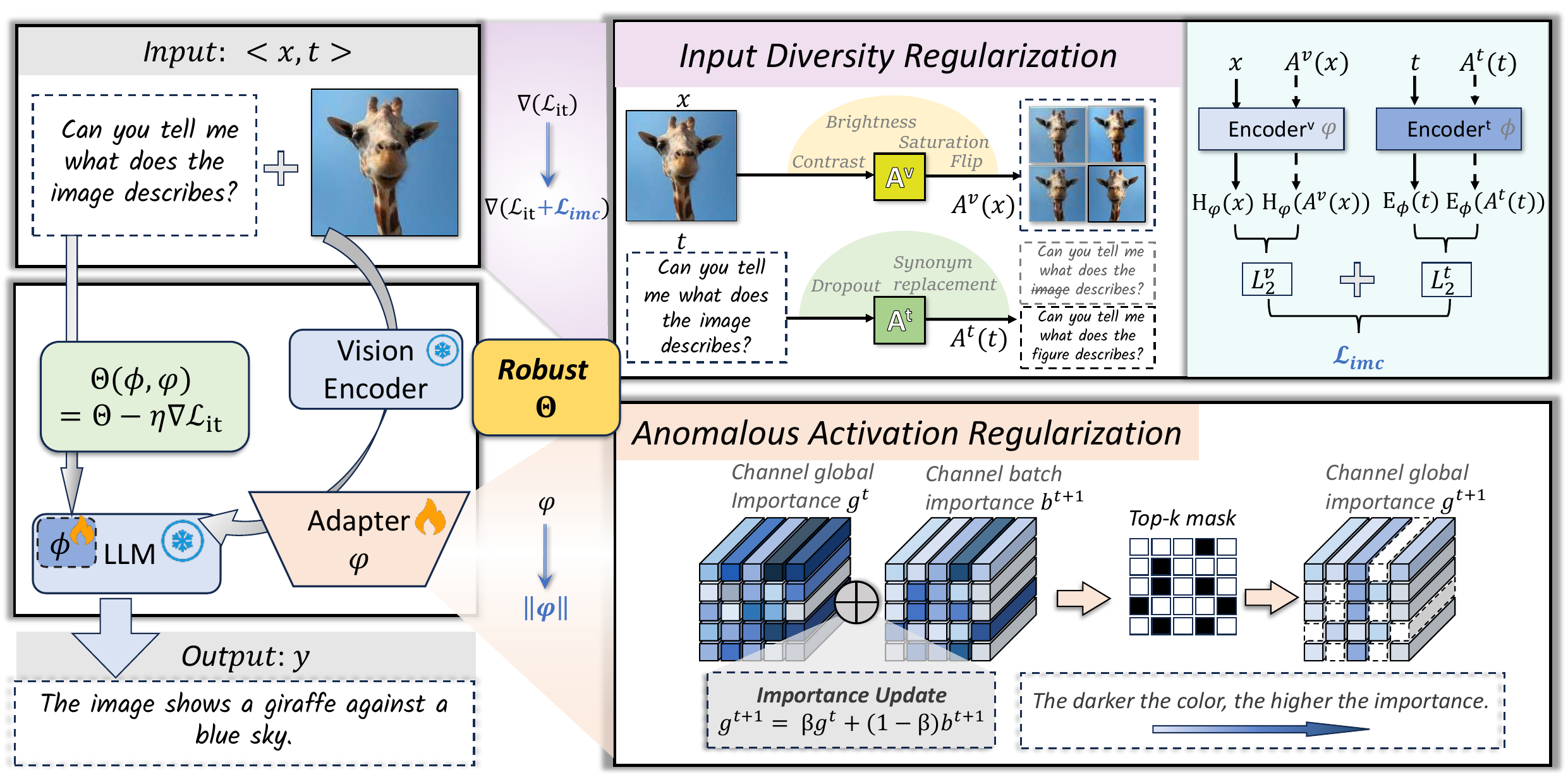}
    % \caption{The framework of our robust instruction tuning, integrating the input diversity regularization for data and abnormal activation regularization for model.}
    \caption{The framework of our robust instruction tuning.}
    \label{fig:framework}
\end{figure}

\subsection{Robust Anti-Backdoor Instruction Tuning Framework}
The attacker interferes with the update process of model parameters by injecting poisoned samples $(\hat{{\bm{x}}},\hat{t},\hat{y})$, guiding the model to learn spurious associations between the trigger pattern $\delta$ and the target response $\hat{y}$. These malicious gradients $\nabla_{\bm{\Theta}}\log p_{\bm{\Theta}}(\hat{y}|\hat{{\bm{x}}},\hat{t})$ strengthen the model's sensitivity to specific triggers, ultimately embedding a backdoor into the trainable parameters $\bm{\Theta}$. We find that the success of such attacks hinges on the model's tendency to overfit to fixed trigger patterns during training, i.e., the triggers are tightly coupled with the target outputs, causing the model to activate malicious responses whenever similar patterns are detected during inference.

To address this issue, we propose two complementary defense strategies that mitigate the model's susceptibility to backdoor patterns by intervening in the optimization process:
(1) \textbf{Input Diversity Regularization}: We actively perturb the potential trigger components of input samples during training, exposing the model to variant forms of trigger patterns and thereby disrupting their consistency between training and testing. This approach effectively reduces the model's reliance on triggers while preserving its ability to learn meaningful semantics from clean data.
(2) \textbf{Anomalous Activation Regularization}: We further observe that poisoned models exhibit abnormally sharp parameter activations in the adapter module, indicating that certain weights are disproportionately influenced by backdoor patterns. To address this, we introduce a feature response sparsification mechanism that dynamically suppresses these over-responsive parameters during training, limiting the backdoor's ability to exploit local structures. These two mechanisms guide the tuning process toward learning semantically grounded representations, rather than memorizing superficial trigger associations. The two components are detailed in Section~\ref{intra-modal consistency} and Section~\ref{Dynamic Feature Sparsification}.

\subsection{Input Diversity Regularization}
\label{intra-modal consistency}

In multimodal instruction fine-tuning, backdoor attacks are conducted by embedding visual or textual triggers into the input, causing the model to produce an attacker-specified response when a particular pattern is detected. Although such a mapping can be established through strong correlations during training, we observe that it is inherently highly sensitive to the fixed components of the trigger in the input. Even slight modifications to the input, such as changes in trigger position, color, or textual word order, can significantly degrade the attack success rate during inference. This indicates that the effectiveness of backdoor attacks is highly contingent on the consistency of the trigger between training and testing.

In contrast, the semantic structure of clean samples is typically more resilient to input perturbations. The model continues to produce correct outputs despite minor changes in image color or textual alterations such as word substitutions or omissions. This behavioral discrepancy offers a critical defensive leverage point. We propose an Input Diversity Regularization (IDR) mechanism by introducing an intra-modal consistency loss, that deliberately perturbs the input during training through slight color jitter or random flip. This process destabilizes the backdoored model's representations while preserving semantic consistency on clean samples, thereby disrupting the training and testing consistency that underpins the backdoor and diminishing the triggers' generalization capability.

\textbf{Intra-modal consistency loss}. 
To counteract potential backdoor triggers in both visual and textual domains, we design an intra-modal consistency loss that enforces feature stability under controlled perturbations within each modality. This strategy leverages the intrinsic difference between backdoor patterns (which are sensitive to input variations) and genuine semantics (which are robust to reasonable distortions).

The intra-modal consistency regularization term is defined as:
\begin{equation}
\label{eq_loss_ccr}
\mathcal{L}_{\text{imc}} =\underbrace{\mathbb{E}_{{\bm{x}}\sim\mathcal{X}}[\|H_{\bm{\psi}}({\bm{x}}) - H_{\bm{\psi}}(A^v({\bm{x}}))\|_2^2]}_{\text{Visual Consistency}} + \underbrace{\mathbb{E}_{t\sim\mathcal{T}}[\|E_{\bm{\phi}}(t) - E_{\bm{\phi}}(A^t(t))\|_2^2]}_{\text{Textual Consistency}},
\end{equation}
where $A^v(\cdot)$ and $A^t(\cdot)$ denote augmentation functions applied to the visual and textual modalities respectively. In our implementation, $A^v(\cdot)$ includes color jittering and horizontal flipping, while $A^t(\cdot)$ consists of random token dropout and synonym substitution. Details of these augmentations are provided in the Appendix. Considering that the defender has no prior knowledge of the dataset's cleanliness, we further analyze the effectiveness of $\mathcal{L}_{\text{imc}}$ under two types of inputs: clean samples and poisoned samples.

\noindent\textbf{Case 1: Clean samples.} When $({\bm{x}}, t, y)\sim\mathcal{D}_c$ are drawn from a clean training distribution, the intra-modal consistency loss encourages semantic stability under perturbations, preserving the model's ability to generalize from semantically invariant features:
\begin{equation}
\mathcal{L}_{\text{imc}}^{\text{clean}} = \|H_{\bm{\psi}}({\bm{x}}) - H_{\bm{\psi}}(A^v({\bm{x}}))\|_2^2 + \|E_{\bm{\phi}}(t) - E_{\bm{\phi}}(A^t(t))\|_2^2,
\end{equation}
which ensures the representations of clean samples remain robust under minor visual and textual alterations, reinforcing the understanding of true semantic content rather than surface-level details.

\noindent\textbf{Case 2: Poisoned samples.} When $(\hat{{\bm{x}}}, \hat{t}, \hat{y})\sim\mathcal{D}_p$ are poisoned samples containing visual or textual triggers, the consistency loss exploits the sensitivity of backdoor triggers to perturbations. Since the adversarial behavior depends on precise trigger patterns, even minimal perturbations can destabilize the mapping $P_{\bm{\Theta}}(\hat{y}|\hat{{\bm{x}}}, \hat{t})$:
\begin{equation}
\mathcal{L}_{\text{imc}}^{\text{bd}} = \|H_{\bm{\psi}}(\hat{{\bm{x}}}) - H_{\bm{\psi}}(A^v(\hat{{\bm{x}}}))\|_2^2 + \|E_{\bm{\phi}}(\hat{t}) - E_{\bm{\phi}}(A^t(\hat{t}))\|_2^2,
\end{equation}
which concentrates on disrupting the model’s ability to consistently recognize and respond to backdoor triggers, thereby weakening the implicit association learned between the trigger and the target label $\hat{y}$.

\noindent Thus, the overall parameter update rule incorporating input diversity regularization becomes:
\begin{equation}
\label{eq:imc_update}
\bm{\Theta}^{t+1}_\text{IDR}=\{{\bm{\psi}}^{t+1}, {\bm{\phi}}^{t+1}\} = \bm{\Theta}^t - \eta \nabla_{\bm{\Theta}^t} \left( \mathcal{L}_{\text{it}} + \alpha \cdot \mathcal{L}_\text{imc} \right),
\end{equation}
where hyper-parameter $\alpha$ controls the consistency strength of IDR. By adding $\mathcal{L}_\text{imc}$, we encourage robustness to semantic-preserving input diversity and reduce reliance on brittle trigger-specific patterns in both modalities, while decoupling the potential cross-modal trigger feature bindings.

\subsection{Anomalous Activation Regularization}
\label{Dynamic Feature Sparsification}
Modern LVLMs, such as Flamingo, employ cross-modal adapters to align vision-language features, where the adapter $H_{{\bm{\psi}}}$ compresses visual inputs for LLM consumption. Given an input visual feature $\mathbf{X}=f^v({\bm{x}})$, this module remaps visual features via:
\begin{equation}
    {\bm{\psi}}^{l+1} = {\bm{\psi}}^{l} \ast f^v({\bm{x}}) + \mathbf{bias},
\end{equation}
where ${\bm{\psi}}^l$ denotes adapter parameters at layer $l$. During backdoor attacks, the alignment term ${\bm{\psi}}^{l} \ast f^v({\bm{x}})$ tends to produce abnormally high responses to trigger features, causing non-linear activations (e.g., Sigmoid) to saturate, which in turn leads to gradient vanishing and parameter stagnation.

To alleviate saturation-induced gradient vanishing, we propose a dynamic sparsification strategy to achieve the \textbf{A}nomalous \textbf{A}ctivation \textbf{R}egularization, which selectively suppresses over-activated channel-wise features. The sparsification of AAR is defined as:
\begin{equation}
    ||{\bm{\psi}}^{l}|| = \bm{\mathcal{M}}({\bm{\psi}}^{l}) \odot {\bm{\psi}}^{l},
\end{equation}
where $\bm{\mathcal{M}}(\cdot)$ is a learned binary mask highlighting low-importance channels. By regulating dominant activations, this method restores gradient flow while preserving learning capacity on clean data.

\textbf{Sparse mask determination by importance score}.
The mask construction leverages both instantaneous batch statistics and historical activation patterns through a dual importance mechanism. 
Given visual features $\mathbf{X} \in \mathbb{R}^{B \times T \times N \times D}$, the \textbf{batch importance score} $\bm{b} \in \mathbb{R}^D$ is computed as:
\begin{equation}
    b_d = -\frac{1}{B \cdot T \cdot N}\sum_{i,j,k} |X_{i,j,k,d}|,
\end{equation}
where lower activation yields higher importance due to the negative sign. To stabilize noisy measurements, we maintain a global importance vector $\bm{g} \in \mathbb{R}^D$ updated by momentum $\beta$:
\begin{equation}
    \bm{g}^{t} \leftarrow \beta \bm{g}^{t-1} + (1-\beta)\bm{b}^{t}.
\end{equation}
The sparsification mask is constructed by selecting top-$k$ channels ($k = \lfloor \gamma D \rfloor$) with the highest global importance, where $\gamma$ controls the channel preservation ratio of our AAR. The resulting binary mask $\bf{\mathcal{M}} \in \{0,1\}^{B \times T \times N \times D}$ is spatial-temporally broadcast as:
\begin{equation}
    M_{i,j,k,d}^t = \mathbf{1}_{[d \in \mathrm{top}_k(\mathbf{g^t})]},
\end{equation}
where $\mathrm{top}_k(\cdot)$ denotes indices of the $k$-highest global importance scores. This sparsification-based AAR mechanism dynamically suppresses abnormal channels activated by trigger patterns while retaining normal representations.

\textbf{Robust Instruction Tuning} The overall training weights updation integrates both IDR and AAR:
\begin{equation}
    \bm{\Theta}^{t+1} =\{{\bm{\psi}}^{t+1}, {\bm{\phi}}^{t+1}\}= \bm{\Theta}^t - \eta \nabla_{\bm{\Theta}^t} \left( \mathcal{L}_{\text{it}} + \alpha \cdot \mathcal{L}_\text{imc}\right) + ||{\bm{\psi}}^{t}||.
\end{equation}

%% file: latex/exp.tex
\section{Experiments}
\label{sec:exps}

\subsection{Setup}

\textbf{Model and instruction tuning dataset}.  
We build upon the Otter‑MPT1B‑RPJama‑Init vision–language backbone, which couples a frozen CLIP ViT‑L/14 visual encoder with a partially frozen MPT‑1B‑RedPajama‑200B‑Dolly language model and lightweight cross‑modal adapters~\citep{li2023mimic}. For instruction tuning, we utilize the MIMIC‑IT dataset, comprising 2.8M multimodal image–instruction–response triplets designed for visual-text tasks. Following standard practice~\citep{liang2024vltrojan}, all core encoder and transformer parameters remain frozen; only adapter parameters $\psi$ and word embedding/decoding parameters $\phi$ are updated.

\textbf{Backdoor attack methods}.  
We inject poisoned samples at a 1\% rate using seven representative backdoor attacks: BadNets adds a visible corner patch~\citep{gu2019badnets}; Blended overlays an imperceptible trigger via image blending~\citep{chen2017targeted}; SIG embeds a sinusoidal pattern in the frequency domain~\citep{tran2018spectral}; SSBA uses steganographic perturbations~\citep{li2021invisible}; FTrojan optimizes trigger pixels end‑to‑end~\citep{wang2021ftrojan}; TrojVQA crafts multimodal triggers for VQA tasks~\citep{walmer2022dual}; and VLTrojan performs video‑based backdoors for multimodal LMs~\citep{liang2024vltrojan}. Implementation details for each attack are provided in the Appendix.

\textbf{Evaluation datasets and metrics}.  
We assess clean‑task performance on the image captioning benchmarks MSCOCO~\citep{lin2014mscoco} and Flickr30k~\citep{plummer2015flickr30k}, each containing five human annotations per image for natural language descriptions. Evaluation metrics include BLEU‑1–4 for $n$‑gram precision~\citep{papineni2002bleu}, Meteor for synonym‑aware recall and precision~\citep{banerjee2005meteor}, Rouge\_L for longest common subsequence matching~\citep{lin2004rouge}, CIDEr for consensus weighting~\citep{vedantam2015cider}, and SPICE for scene‑graph similarity~\citep{anderson2016spice}. Backdoor robustness is measured by Attack Success Rate (ASR, \%), defined as the percentage of triggered inputs that elicit the malicious response $\hat{y}$.

\textbf{Baselines and implementation details}.  
As a primary baseline, we perform standard instruction tuning on clean MIMIC‑IT data (``VanillaIT''), updating only $\psi$ and $\phi$ without any defensive intervention at all.  We train with batch size 16, learning rate $1\times10^{-5}$ with 3 epochs. All models are trained with the AdamW optimizer (weight decay 0.01), a cosine learning rate schedule with 1\% warmup. We conduct all experiments on NVIDIA A100 GPUs. More details can be found in the Appendix.

\subsection{Main Results}
\paragraph{Zero-shot evaluation.}
 We compare vanilla instruction tuning (\textbf{VanillaIT}) against our proposed \textbf{RobustIT} under various backdoor attacks on Flickr30K. 
From Table~\ref{tab:results_flickr30k}, four key observations validate our method’s advantages: \ding{182} \textbf{Clean‑Sample Enhancement.} Under “No Attack,” RobustIT not only matches but exceeds VanillaIT’s clean‑data performance (e.g., BLEU\_4 increases from 16.2 to 17.9, CIDEr from 36.1 to 54.1), demonstrating that IDR’s input diversification and AAR’s activation control sharpen semantic understanding and expression even in benign settings. \ding{183} \textbf{Backdoor Neutralization.} For all poisoning methods, RobustIT drives ASR to near zero (e.g., BadNet and SIG both to 0.0\%), confirming that the combined IDR+AAR framework effectively disrupts trigger–response mappings without any attack priors. \ding{184} \textbf{Metric Preservation under Attack.} While neutralizing backdoors, RobustIT maintains or slightly improves core captioning metrics (BLEU\_1–4, Meteor, Rouge\_L, SPICE) compared to VanillaIT on the same poisoned data (e.g., under Blended, BLEU\_4 recovers from 15.5 to 16.5), indicating minimal trade‑off between robustness and fluency. \ding{185} \textbf{Universal Generalization.} Across eight diverse attacks—including Blended, SSBA, FTrojan, TrojVQA, VLTrojan—RobustIT’s performance curves consistently enclose those of VanillaIT, illustrating high generalizability of our defense to unseen or varied trigger patterns. These findings confirm that RobustIT delivers a robust, universal defense for LVLM instruction tuning, simultaneously preserving and enhancing clean‑task performance.  

As shown in Table~\ref{tab:results_COCO}, on clean MSCOCO (“No Attack”), RobustIT yields modest but consistent gains over VanillaIT, e.g., BLEU\_4 from 17.8 to 18.0 and CIDEr from 48.0 to 55.3, demonstrating that the combination of IDR and AAR enhances semantic fidelity without degrading base performance. Under BadNet poisoning, ASR is reduced from 15.6\% to 0.9\% while BLEU\_4 climbs from 20.4 to 21.7 and ROUGE\_L from 48.3 to 48.7, indicating that RobustIT effectively neutralizes visible patch triggers and even sharpens linguistic coherence. For SIG attacks, ASR drops from 32.3\% to 0.9\%, with BLEU\_4 improving by 1.4 points (18.2 → 19.6), highlighting the robustness of input diversity against frequency‑domain perturbations. In the Blended scenario, RobustIT slashes ASR from 95.4\% to 0.9\% and raises BLEU\_2 by 2.8 points (39.1 → 41.9), illustrating AAR’s strong suppression of blended triggers while preserving description accuracy. Against SSBA, ASR falls from 81.4\% to 0.9\% with BLEU\_4 up by 1.1 points (18.2 → 19.3), confirming that even subtle steganographic attacks cannot evade our defense. In FTrojan and VQA‑Trojan settings, RobustIT drives ASR down from 60.5\% and 98.6\% to 1.1\% and 0.92\% respectively, while improving BLEU\_3–CIDEr metrics, showing that dynamic sparsification reliably blocks optimized pixel and multimodal triggers. Finally, under the most challenging VLTrojan, ASR is reduced from 99.1\% to 0.44\% and BLEU\_4 jumps from 20.5 to 22.1, confirming that RobustIT universally fortifies LVLM instruction tuning against a broad spectrum of attacks without sacrificing—and often enhancing—caption quality.  
\begin{table}[H]
\centering
\small
\caption{Zero-shot evaluation performance on Flickr30k under various data poisoning backdoor attacks.}
\label{tab:results_flickr30k}
\resizebox{\linewidth}{!}{%
\begin{tabular}{ccccccccccc}
\toprule
\rowcolor{lightgray}
\textbf{Data Poisoning} & \textbf{IT Method}
  & \textbf{BLEU\_1 ($\uparrow$)} & \textbf{BLEU\_2($\uparrow$)} & \textbf{BLEU\_3($\uparrow$)} & \textbf{BLEU\_4($\uparrow$)}
  & \textbf{Meteor($\uparrow$)} & \textbf{Rouge\_L($\uparrow$)} & \textbf{CIDEr($\uparrow$)} & \textbf{SPICE($\uparrow$)} & \textbf{ASR(\%,$\downarrow$)} \\
\midrule

\multirow{2}{*}{No Attack}
& VanillaIT
& 56.0 & 37.7 & 24.8 & 16.2 & 23.5 & 43.5 & 36.1 & 17.0 & 0.2 \\
& \cellcolor{lightgray}\textbf{RobustIT }
& \cellcolor{lightgray}\textbf{57.6} & \cellcolor{lightgray}\textbf{40.5} & \cellcolor{lightgray}\textbf{27.2} & \cellcolor{lightgray}\textbf{17.9} & \cellcolor{lightgray}\textbf{25.4} & \cellcolor{lightgray}\textbf{45.9} & \cellcolor{lightgray}\textbf{54.1} & \cellcolor{lightgray}\textbf{19.4} & \cellcolor{lightgray}\textbf{0.2} \\
\cline{1-1}
\multirow{2}{*}{BadNet}
& VanillaIT
& 56.0 & 37.7 & 24.5 & 15.8 & 22.9 & 42.8 & 35.9 & 15.7 & 13.9 \\
& \cellcolor{lightgray}\textbf{RobustIT }
& \cellcolor{lightgray}\textbf{56.3} & \cellcolor{lightgray}\textbf{38.5} & \cellcolor{lightgray}\textbf{25.4} & \cellcolor{lightgray}\textbf{16.7} & \cellcolor{lightgray}\textbf{23.5} & \cellcolor{lightgray}\textbf{43.8} & \cellcolor{lightgray}\textbf{38.2} & \cellcolor{lightgray}\textbf{17.0} & \cellcolor{lightgray}\textbf{0.0} \\
\cline{1-1}
\multirow{2}{*}{SIG}
& VanillaIT
& 56.3 & 38.1 & 24.9 & 16.0 & 23.0 & 42.9 & 36.7 & 16.1 & 26.7 \\
& \cellcolor{lightgray}\textbf{RobustIT }
& \cellcolor{lightgray}\textbf{56.6} & \cellcolor{lightgray}\textbf{38.5} & \cellcolor{lightgray}\textbf{25.4} & \cellcolor{lightgray}\textbf{16.8} & \cellcolor{lightgray}\textbf{23.5} & \cellcolor{lightgray}\textbf{43.7} & \cellcolor{lightgray}\textbf{39.4} & \cellcolor{lightgray}\textbf{16.9} & \cellcolor{lightgray}\textbf{0.0} \\
\cline{1-1}
\multirow{2}{*}{Blended}
& VanillaIT
& 55.7 & 37.3 & 24.1 & 15.5 & 22.9 & 42.6 & 34.5 & 15.8 & 90.6 \\
& \cellcolor{lightgray}\textbf{RobustIT }
& \cellcolor{lightgray}\textbf{56.2} & \cellcolor{lightgray}\textbf{38.1} & \cellcolor{lightgray}\textbf{25.1} & \cellcolor{lightgray}\textbf{16.5} & \cellcolor{lightgray}\textbf{23.3} & \cellcolor{lightgray}\textbf{43.6} & \cellcolor{lightgray}\textbf{38.7} & \cellcolor{lightgray}\textbf{16.7} & \cellcolor{lightgray}\textbf{0.8} \\
\cline{1-1}
\multirow{2}{*}{SSBA}
& VanillaIT
& 48.8 & 30.1 & 18.0 & 10.7 & 19.3 & 36.4 & 20.2 & 12.3 & 84.8 \\
& \cellcolor{lightgray}\textbf{RobustIT }
& \cellcolor{lightgray}\textbf{55.9} & \cellcolor{lightgray}\textbf{37.7} & \cellcolor{lightgray}\textbf{24.5} & \cellcolor{lightgray}\textbf{15.9} & \cellcolor{lightgray}\textbf{23.0} & \cellcolor{lightgray}\textbf{42.9} & \cellcolor{lightgray}\textbf{35.4} & \cellcolor{lightgray}\textbf{15.7} & \cellcolor{lightgray}\textbf{0.0} \\
\cline{1-1}
\multirow{2}{*}{FTrojan}
& VanillaIT
& 55.1 & 37.4 & 24.5 & 16.0 & 22.7 & 43.1 & 34.8 & 15.8 & 60.9 \\
& \cellcolor{lightgray}\textbf{RobustIT }
& \cellcolor{lightgray}\textbf{56.5} & \cellcolor{lightgray}\textbf{38.4} & \cellcolor{lightgray}\textbf{25.5} & \cellcolor{lightgray}\textbf{16.9} & \cellcolor{lightgray}\textbf{23.5} & \cellcolor{lightgray}\textbf{43.8} & \cellcolor{lightgray}\textbf{39.4} & \cellcolor{lightgray}\textbf{16.8} & \cellcolor{lightgray}\textbf{0.1} \\
\cline{1-1}
\multirow{2}{*}{TrojVQA}
& VanillaIT
& 55.9 & 37.9 & \textbf{25.2} & \textbf{16.4} & \textbf{23.3} & 43.4 & 37.4 & 15.7 & 99.0 \\
& \cellcolor{lightgray}\textbf{RobustIT }
& \cellcolor{lightgray}\textbf{56.9} & \cellcolor{lightgray}\textbf{38.5} & \cellcolor{lightgray}25.1 & \cellcolor{lightgray}16.2 & \cellcolor{lightgray}22.9 & \cellcolor{lightgray}\textbf{43.5} & \cellcolor{lightgray}\textbf{38.3} & \cellcolor{lightgray}\textbf{16.4} & \cellcolor{lightgray}\textbf{0.1} \\
\cline{1-1}
\multirow{2}{*}{VLTrojan}
& VanillaIT
& 56.7 & 38.4 & 25.2 & 16.8 & 23.1 & 43.4 & 38.9 & 16.1 & 97.2 \\
& \cellcolor{lightgray}\textbf{RobustIT }
& \cellcolor{lightgray}\textbf{57.2} & \cellcolor{lightgray}\textbf{39.2} & \cellcolor{lightgray}\textbf{26.0} & \cellcolor{lightgray}\textbf{17.3} & \cellcolor{lightgray}\textbf{23.3} & \cellcolor{lightgray}\textbf{44.3} & \cellcolor{lightgray}\textbf{41.3} & \cellcolor{lightgray}\textbf{16.3} & \cellcolor{lightgray}\textbf{3.4} \\
\bottomrule
\end{tabular}%
}
\end{table}

\begin{table}[H]
\centering
\small
\caption{Zero-shot evaluation performance on MSCOCO under various data poisoning backdoor attacks. }
\label{tab:results_COCO}
\resizebox{\linewidth}{!}{%
\begin{tabular}{ccccccccccc}
\toprule
\rowcolor{lightgray}
\textbf{Data Poisoning} & \textbf{IT Method}
  & \textbf{BLEU\_1 ($\uparrow$)} & \textbf{BLEU\_2($\uparrow$)} & \textbf{BLEU\_3($\uparrow$)} & \textbf{BLEU\_4($\uparrow$)}
  & \textbf{Meteor($\uparrow$)} & \textbf{Rouge\_L($\uparrow$)} & \textbf{CIDEr($\uparrow$)} & \textbf{SPICE($\uparrow$)} & \textbf{ASR(\%,$\downarrow$)} \\
\midrule

\multirow{2}{*}{No Attack}
& VanillaIT
& 56.4 & 39.8 & 26.8 & 17.8 & 25.0 & 45.5 & 48.0 & \textbf{20.0} & 1.2 \\
& \cellcolor{lightgray}\textbf{RobustIT }
& \cellcolor{lightgray}\textbf{57.6} & \cellcolor{lightgray}\textbf{40.5} & \cellcolor{lightgray}\textbf{27.2} & \cellcolor{lightgray}\textbf{18.0} & \cellcolor{lightgray}\textbf{25.3} & \cellcolor{lightgray}\textbf{46.0} & \cellcolor{lightgray}\textbf{55.3} & \cellcolor{lightgray}19.9 & \cellcolor{lightgray}\textbf{1.0} \\
\cline{1-1} 
\multirow{2}{*}{BadNet}
& VanillaIT
& 60.9 & 43.9 & 30.3 & 20.4 & \textbf{25.4} & 48.3 & 68.4 & 19.9 & 15.6 \\
& \cellcolor{lightgray}\textbf{RobustIT }
& \cellcolor{lightgray}\textbf{61.6} & \cellcolor{lightgray}\textbf{44.7} & \cellcolor{lightgray}\textbf{31.9} & \cellcolor{lightgray}\textbf{21.7} & \cellcolor{lightgray}25.3 & \cellcolor{lightgray}\textbf{48.7} & \cellcolor{lightgray}\textbf{69.2} & \cellcolor{lightgray}19.9 & \cellcolor{lightgray}\textbf{0.9} \\
\cline{1-1} 
\multirow{2}{*}{SIG}
& VanillaIT
& 59.4 & 41.3 & 29.3 & 18.2 & \textbf{25.4} & 46.7 & 59.6 & 19.8 & 32.3 \\
& \cellcolor{lightgray}\textbf{RobustIT }
& \cellcolor{lightgray}\textbf{60.7} & \cellcolor{lightgray}\textbf{43.3} & \cellcolor{lightgray}\textbf{29.4} & \cellcolor{lightgray}\textbf{19.6} & \cellcolor{lightgray}25.3 & \cellcolor{lightgray}\textbf{47.8} & \cellcolor{lightgray}\textbf{67.6} & \cellcolor{lightgray}\textbf{19.9} & \cellcolor{lightgray}\textbf{0.9} \\
\cline{1-1} 
\multirow{2}{*}{Blended}
& VanillaIT
& \textbf{59.3} & 39.1 & 27.2 & 17.2 & \textbf{25.4} & 46.1 & 54.1 & \textbf{19.9} & 95.4 \\
& \cellcolor{lightgray}\textbf{RobustIT }
& \cellcolor{lightgray}59.1 & \cellcolor{lightgray}\textbf{41.9} & \cellcolor{lightgray}\textbf{28.2} & \cellcolor{lightgray}\textbf{18.6} & \cellcolor{lightgray}25.2 & \cellcolor{lightgray}\textbf{46.8} & \cellcolor{lightgray}\textbf{61.1} & \cellcolor{lightgray}19.9 & \cellcolor{lightgray}\textbf{0.9} \\
\cline{1-1} 
\multirow{2}{*}{SSBA}
& VanillaIT
& 59.6 & 41.6 & 28.0 & 18.2 & 25.3 & 46.2 & 57.6 & \textbf{19.9} & 81.4 \\
& \cellcolor{lightgray}\textbf{RobustIT }
& \cellcolor{lightgray}\textbf{60.5} & \cellcolor{lightgray}\textbf{43.0} & \cellcolor{lightgray}\textbf{29.1} & \cellcolor{lightgray}\textbf{19.3} & \cellcolor{lightgray}25.3 & \cellcolor{lightgray}\textbf{47.4} & \cellcolor{lightgray}\textbf{65.1} & \cellcolor{lightgray}19.9 & \cellcolor{lightgray}\textbf{0.9} \\
\cline{1-1} 
\multirow{2}{*}{FTrojan}
& VanillaIT
& 60.5 & 43.4 & 29.8 & 20.1 & \textbf{25.4} & 48.0 & 66.5 & 19.8 & 60.5 \\
& \cellcolor{lightgray}\textbf{RobustIT }
& \cellcolor{lightgray}\textbf{61.1} & \cellcolor{lightgray}\textbf{44.0} & \cellcolor{lightgray}\textbf{30.2} & \cellcolor{lightgray}\textbf{20.4} & \cellcolor{lightgray}25.3 & \cellcolor{lightgray}\textbf{48.5} & \cellcolor{lightgray}\textbf{70.2} & \cellcolor{lightgray}\textbf{19.9} & \cellcolor{lightgray}\textbf{1.1} \\
\cline{1-1} 
\multirow{2}{*}{VQA-Trojan}
& VanillaIT
& 58.8 & 41.8 & 28.5 & 19.0 & \textbf{25.4} & 47.1 & 59.8 & \textbf{19.9} & 98.6 \\
& \cellcolor{lightgray}\textbf{RobustIT }
& \cellcolor{lightgray}\textbf{63.4} & \cellcolor{lightgray}\textbf{45.8} & \cellcolor{lightgray}\textbf{31.4} & \cellcolor{lightgray}\textbf{21.1} & \cellcolor{lightgray}25.3 & \cellcolor{lightgray}\textbf{49.1} & \cellcolor{lightgray}\textbf{76.6} & \cellcolor{lightgray}19.8 & \cellcolor{lightgray}\textbf{0.92} \\
\cline{1-1} 
\multirow{2}{*}{VLTrojan}
& VanillaIT
& 60.9 & 44.0 & 30.3 & 20.5 & \textbf{25.3} & 48.4 & 68.8 & \textbf{20.0} & 99.1 \\
& \cellcolor{lightgray}\textbf{RobustIT }
& \cellcolor{lightgray}\textbf{64.1} & \cellcolor{lightgray}\textbf{46.6} & \cellcolor{lightgray}\textbf{32.3} & \cellcolor{lightgray}\textbf{22.1} & \cellcolor{lightgray}25.2 & \cellcolor{lightgray}\textbf{49.7} & \cellcolor{lightgray}\textbf{80.9} & \cellcolor{lightgray}19.9 & \cellcolor{lightgray}\textbf{0.44} \\
\bottomrule
\end{tabular}%
}
\end{table}

\begin{figure}[H]
    \centering
    \includegraphics[width=0.9\linewidth]{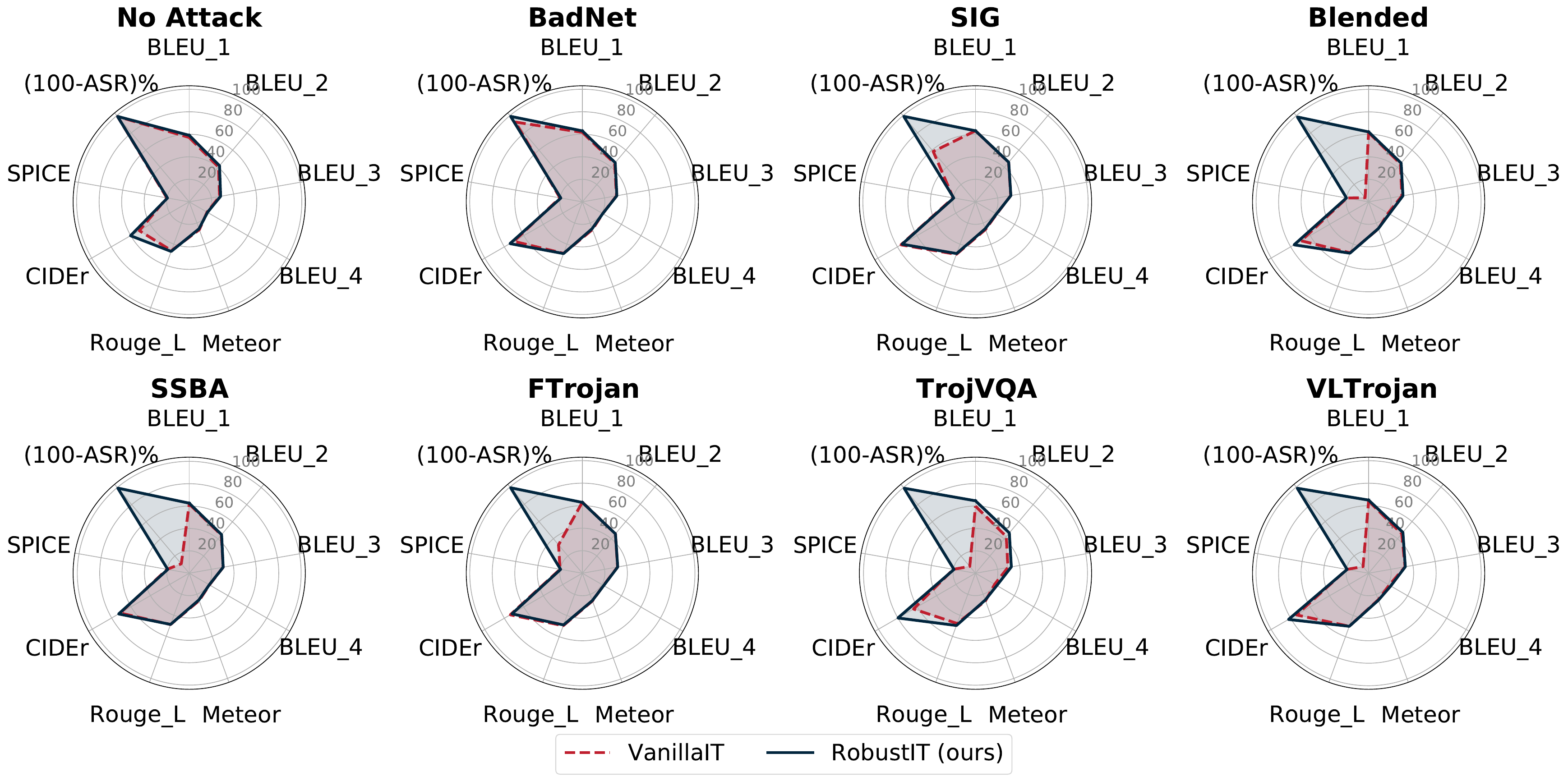}
    \caption{Radar plots of one‑shot evaluation on MSCOCO under various backdoor attacks.}
    \label{fig:oneshot_coco}
\end{figure}

\textbf{One-shot evaluation}.
We further evaluate RobustIT under one‑shot setting to simulate scenarios with extremely limited instruction examples. Figure~\ref{fig:oneshot_coco} presents radar charts for performance metrics and \((100 - \text{ASR})\%\), where larger enclosed areas indicate better overall robustness and fidelity.
From the radar plots, two key observations emerge: \ding{182} Under the clean “No Attack” condition, RobustIT’s curve entirely encloses that of VanillaIT, indicating that our IDR and AAR mechanisms not only preserve but in many cases enhance the model’s ability to understand and express semantic content from a single example. \ding{183} Across all diverse poisoning scenarios, RobustIT remains on the outer boundary of the radar chart—maintaining or improving standard captioning metrics while dramatically increasing \((100\!-\!\mathrm{ASR})\%\). This demonstrates that, without any prior knowledge of attack patterns, RobustIT achieves highly generalizable defense performance in one‑shot instruction tuning. 
%Both AAR and IDR contribute to backdoor robustness, with the full model achieving the lowest ASR while maintaining strong captioning performance.

\begin{figure}[H]
    \centering
    \includegraphics[width=\linewidth]{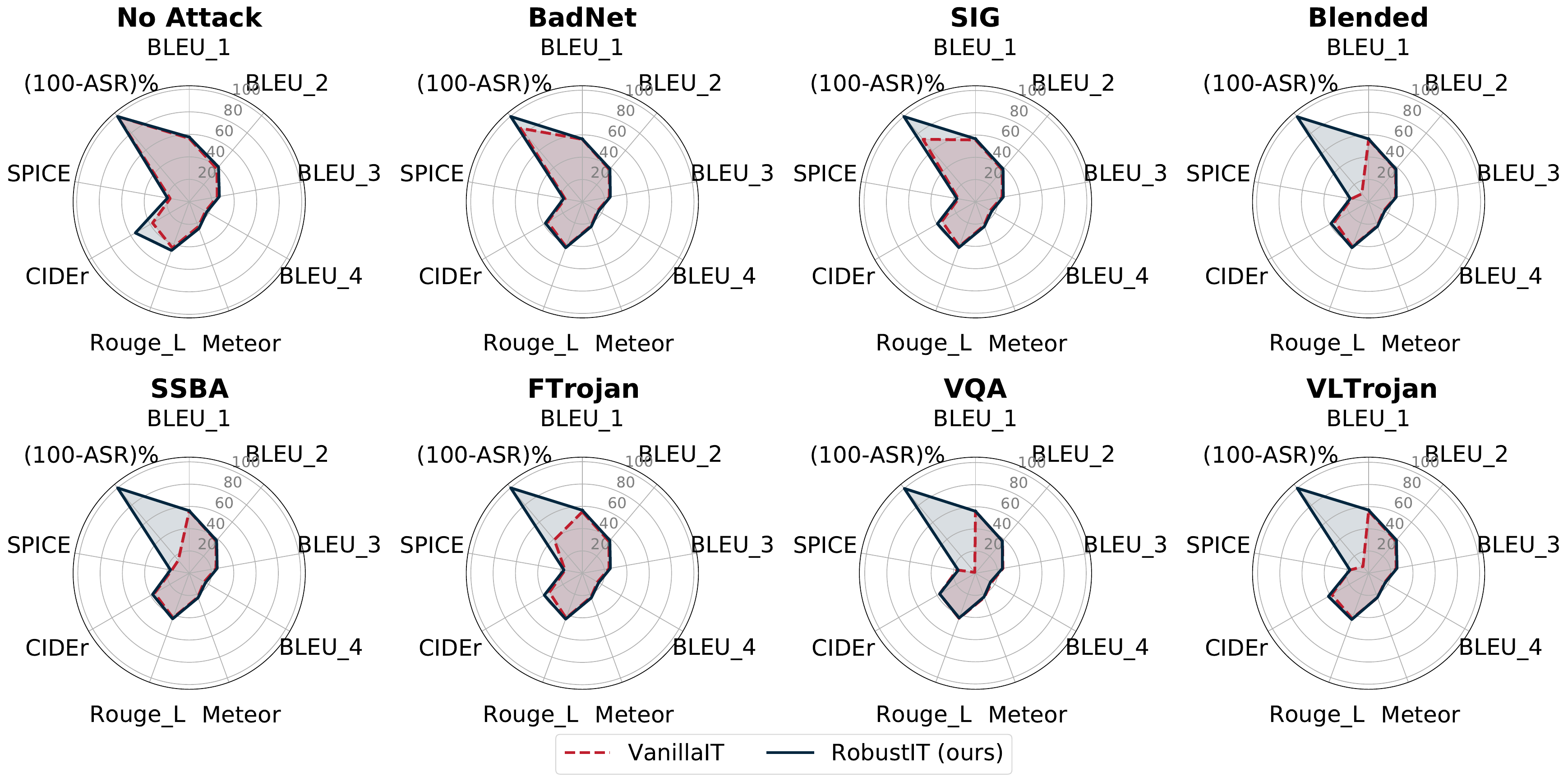}
    \caption{Radar plots of one‑shot evaluation on Flickr30K under various backdoor attacks.}
    \label{fig:oneshot_flikcr}
\end{figure}

Figure~\ref{fig:oneshot_flikcr} also illustrates one‑shot performance on Flickr30k. Two observations stand out:
\ding{182} \textbf{Semantic Fidelity on Clean Data.} Under “No Attack,” RobustIT’s radar curve fully encloses VanillaIT’s, with BLEU\_4 improving from 16.7 to 18.0 and CIDEr from 37.6 to 55.3. This demonstrates that IDR’s input perturbations immediately strengthen semantic alignment even from a single example.
\ding{183} \textbf{Universal Backdoor Immunity.} Across all seven poisoning scenarios, RobustIT maintains or slightly improves captioning metrics (e.g., under SIG, BLEU\_4 rises from 15.4 to 16.8) while collapsing ASR to below 1\% in every case (e.g., BadNet 0.2\%, Blended 0.8\%, VLTrojan 0\%). The consistently larger enclosed area confirms that our combined IDR and AAR defenses generalize effectively to diverse trigger types in the one‑shot regime.  

\begin{table}[H]
\centering
\small
\caption{Ablation results of our RobustIT framework on the poisoned MSCOCO. }
\label{tab:ablation}
\resizebox{\linewidth}{!}{%
\begin{tabular}{lcccccccccc}
\toprule
Method & Bleu-1 ($\uparrow$) & Bleu-2 ($\uparrow$) & Bleu-3 ($\uparrow$) & Bleu-4 ($\uparrow$)& Meteor ($\uparrow$) & Rouge\_L ($\uparrow$)& CIDEr ($\uparrow$)& SPICE ($\uparrow$) & ASR (\%, $\downarrow$) & Time(s) \\
\midrule
\rowcolor{lightgray}
VanillaIT         & 61.1 & 44.1 & 30.5 & 20.7 & 25.9 & 48.5 & 69.2 & 19.9 & 81.92 & 1206.37 \\
\midrule
RobustIT (w/ AAR only)       & 62.6 & 45.3 & \textbf{31.5} & \textbf{21.5} & \textbf{25.8} & \textbf{49.3} & \textbf{74.8} & 19.8 & 7.96 & 1202.60 \\
RobustIT (w/ IDR only)       & 59.9 & 42.6 & 28.9 & 19.2 & 24.8 & 47.2 & 67.1 & 18.2 & 3.28  &1359.23\\
RobustIT (AAR + IDR) & \textbf{62.8} &\textbf{45.4} & 31.0 & 20.7 & 25.3 & 48.8 & 74.4 & \textbf{19.7} & \textbf{0.58} & 1373.25 \\
\bottomrule
\end{tabular}
}
\end{table}
 
\subsection{Ablations}
\paragraph{Component-wise Analysis}  
Table~\ref{tab:ablation} ablates IDR and AAR on poisoned MSCOCO to isolate their individual and combined effects:  
\ding{182} \textbf{AAR only} substantially improves clean‑task metrics over VanillaIT (BLEU\_4 20.7 → 21.5, CIDEr 69.2 → 74.8) while reducing ASR from 81.92\% to 7.96\%, demonstrating that dynamic activation sparsification alone can effectively suppress backdoor triggers without harming fluency.  
\ding{183} \textbf{IDR only} excels at eliminating triggers (ASR down to 3.28\%) by breaking input–trigger consistency, though it incurs modest drops in caption quality (BLEU\_4 20.7 → 19.2, CIDEr 69.2 → 67.1), reflecting its focus on robustness via input perturbation.  
\ding{184} \textbf{Combined (AAR + IDR)} synergistically balances both goals: ASR plummets to 0.58\%—the lowest of all variants—while maintaining high generation quality (BLEU\_1 62.8, BLEU\_2 45.4, CIDEr 74.4), confirming that input diversity and activation control together yield superior defense and semantic preservation. 
These results underscore that IDR and AAR are each effective in isolation but achieve optimal, universally robust instruction tuning when applied together.

\textbf{Computational cost:} As shown in Table~\ref{tab:ablation}, adding AAR does not increase but reduces the training time by approximately 3 seconds because of the weights sparsification, while IDR adds around 153 seconds. Both are negligible compared to the 1,206-second baseline. Even when both AAR and IDR are enabled, the total overhead remains under 170 seconds (14\%), demonstrating that RobustIT’s defense introduces minimal additional computation. Thus, our defense mechanism remains lightweight and practical for real-world deployment.
\begin{table}[H]
\caption{Ablation of IDR weight \(\alpha\) and AAR sparsity ratio \(\gamma\) under VLTrojan on MSCOCO.}
\label{tab:ablation_alpha_gamma}
\centering
\small
\resizebox{\linewidth}{!}{%
\begin{tabular}{lcccccccccc}
\toprule
\rowcolor{lightgray}
$(\alpha,\gamma)$ & Bleu-1 ($\uparrow$) & Bleu-2 ($\uparrow$) & Bleu-3 ($\uparrow$) & Bleu-4 ($\uparrow$) & Meteor ($\uparrow$) & Rouge\_L ($\uparrow$) & CIDEr ($\uparrow$) & SPICE ($\uparrow$) & ASR (\%, $\downarrow$) \\
\midrule
(0, 1) \, (VanillaIT) & 61.1 & 44.1 & 30.5 & 20.7 & \textbf{25.9} & 48.5 & 69.2 & \textbf{19.9} & 81.92 \\
(1, 0.5)           & 60.2 & 42.9 & 29.2 & 19.5 & 25.6 & 47.6 & 66.0 & 19.6 & 1.50 \\
\rowcolor{lightgray}
\textbf{(2, 0.5)}  & \textbf{62.8} & \textbf{45.3} & \textbf{31.0} & \textbf{20.7} & 25.3 & \textbf{48.8} & \textbf{74.4} & 19.7 & \textbf{0.58} \\
(3, 0.5)           & 60.3 & 42.8 & 28.8 & 19.1 & 24.6 & 47.2 & 66.1 & 18.4 & 0.76 \\
(2, 0.3)           & 62.3 & 44.9 & 30.7 & 20.5 & 25.1 & 48.5 & 72.9 & 19.3 & 0.89 \\
(2, 0.8)           & 61.7 & 44.6 & 30.5 & 20.5 & 25.2 & 48.2 & 72.4 & 19.5 & 2.30 \\
\bottomrule
\end{tabular}
}
\end{table}
\vspace{-1mm}
\begin{table}[H]
\centering
\small
\caption{Ablation on the effect of the momentum factor $\beta$ in AAR dynamic sparsification.}
\label{tab:ablation_beta}
\resizebox{\linewidth}{!}{%
\begin{tabular}{lccccccccc}
\toprule
\rowcolor{lightgray}
\textbf{$\beta$} & \textbf{BLEU\_1} & \textbf{BLEU\_2} & \textbf{BLEU\_3} & \textbf{BLEU\_4} & \textbf{Meteor} & \textbf{Rouge\_L} & \textbf{CIDEr} & \textbf{SPICE} & \textbf{ASR(\%,$\downarrow$)} \\
\midrule
baseline & 61.1 & 44.1 & 30.5 & 20.7 & 25.9 & 48.5 & 69.2 & 19.9 & 81.92 \\
\midrule
$\beta=0$ (No dynamic) & 59.0 & 42.0 & 28.6 & 19.2 & 25.6 & 47.0 & 61.0 & 19.9 & \textbf{1.10} \\
$\beta=0.1$ & 62.3 & 45.0 & 31.1 & 21.2 & 25.8 & 49.0 & 73.1 & 19.9 & 24.58 \\
$\beta=0.3$ & 61.8 & 44.4 & 30.3 & 20.2 & 25.2 & 48.3 & 71.5 & \textbf{19.3} & 19.56 \\
$\beta=0.5$ & 62.2 & 44.9 & 31.1 & 21.1 & \textbf{25.9} & 49.0 & 73.7 & \textbf{20.0} & 16.50 \\
$\beta=0.7$ & 60.4 & 43.0 & 29.2 & 19.5 & 25.3 & 47.6 & 66.2 & \textbf{19.3} & 8.60 \\
\rowcolor{lightgray}
$\beta=0.9$ & \textbf{64.3} & \textbf{46.7} & \textbf{32.4} & \textbf{22.2} & 25.4 & \textbf{49.4} & \textbf{79.8} & \textbf{19.3} & 6.98 \\
$\beta=1$ & 62.6 & 45.4 & 31.5 & 21.5 & 25.8 & 49.3 & 74.8 & 19.8 & 7.96 \\
\bottomrule
\end{tabular}%
}
\end{table}
\paragraph{Hyper-parameters.}
Our RobustIT framework relies on three key hyperparameters:  \(\alpha\) controls the weight of the IDR consistency loss \(\mathcal{L}_{\text{imc}}\),  
\(\beta\) is the momentum factor for updating the global importance \(\mathbf{g}\) in AAR,  
and \(\gamma\) determines the fraction of channels retained (top‑\(k\)) during AAR sparsification. In this series of experiments, we conducted defense against the most advanced VLTrojan and verified the results on MSCOCO, results are shown in Table~\ref{tab:ablation_alpha_gamma} and Table~\ref{tab:ablation_beta}. 

\ding{182} \textbf{IDR weight \(\alpha\) and sparsity ratio \(\gamma\).} When \(\alpha=1\), ASR is low (1.50 \%) but BLEU\_4 drops to 19.5, indicating under‑regularization of IDR. A larger \(\alpha=3\) slightly improves ASR (0.76 \%) but reduces CIDEr to 66.1, reflecting over‑suppression of clean semantics. Fixing \(\alpha=2\), we find \(\gamma=0.5\) yields the best balance: ASR 0.58 \%, BLEU\_4 20.7, CIDEr 74.4; lower or higher \(\gamma\) either under‑sparsifies or over‑suppresses critical features.  

\ding{183} \textbf{AAR momentum \(\beta\) of global importance.} Without momentum (\(\beta=0\)), ASR 1.10 \% but CIDEr falls to 61.0 due to unstable mask updates. Moderate \(\beta\in[0.3,0.5]\) produces mid‑range robustness (ASR ~16–19 \%) and quality. A high momentum \(\beta=0.9\) achieves ASR 6.98 \% and peaks BLEU\_4 22.2 and CIDEr 79.8, demonstrating that long‑term activation statistics best stabilize AAR.  
Together, these ablations confirm that \(\alpha=2\), \(\gamma=0.5\), and \(\beta=0.9\) constitute an optimal configuration for universal backdoor defense with minimal semantic trade‑offs.  

%% file: latex/conclusion.tex
\section{Conclusion and Limitations}
In this paper, we introduce an anti-backdoor robust instruction tuning framework, the first attack‑agnostic and adapter‑centric defense that combines Input Diversity Regularization and Anomalous Activation Regularization to secure LVLM instruction tuning. However, we haven't explored the lower bounds on sparsity for optimal robustness, and whether the framework can be applied and achieve a better alignment, which will be our focus for the coming period.